# Probing Single Vacancies in Black Phosphorus at the Atomic Level

Brian Kiraly, Nadine Hauptmann, Alexander N. Rudenko, Mikhail I. Katsnelson, Alexander A. Khajetoorians[*]

Institute for Molecules and Materials, Radboud University, 6525 AJ Nijmegen, Netherlands

[*] Correspondence to: a.khajetoorians@science.ru.nl

Utilizing a combination of low-temperature scanning tunneling microscopy/spectroscopy (STM/STS) and electronic structure calculations, we characterize the structural and electronic properties of single atomic vacancies within several monolayers of the surface of black phosphorus. We illustrate, with experimental analysis and tight-binding calculations, that we can depth profile these vacancies and assign them to specific sublattices within the unit cell. Measurements reveal that the single vacancies exhibit strongly anisotropic and highly delocalized charge density, laterally extended up to 20 atomic unit cells. The vacancies are then studied with STS, which reveals in-gap resonance states near the valence band edge and a strong *p*-doping of the bulk black phosphorus crystal. Finally, quasiparticle interference generated near these vacancies enables the direct visualization of the anisotropic band structure of black phosphorus.

TOC Graphic

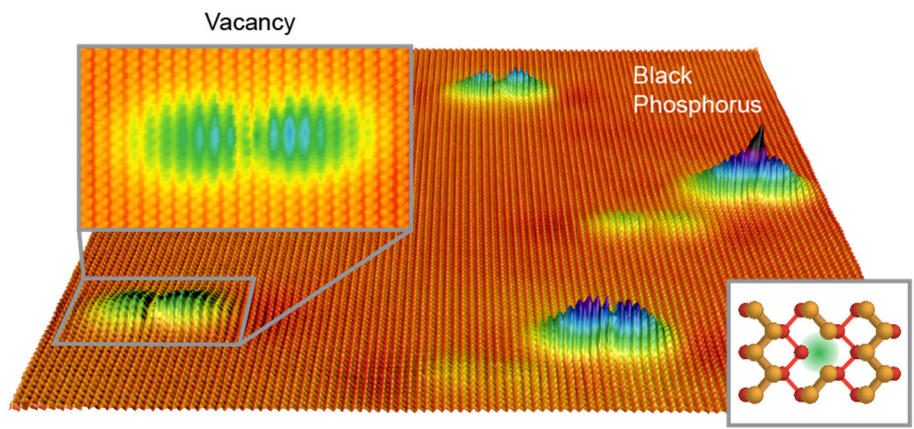

Black phosphorus (BP) is a layered allotrope of phosphorus which crystallizes into an orthorhombic lattice [1]. Weakly bound phosphorene layers are vertically stacked in an arrangement analogous to graphite; in contrast, however, each phosphorene layer possesses two planes of phosphorus atoms bonded though *sp³* orbitals. This leaves two electrons per phosphorus atom paired, yet unbonded [2,3]. The 1D buckling and lone-pair electronic configuration result in an electronic band gap for bulk black phosphorus of 0.3 eV [1], which is predicted to be dramatically enhanced in the single layer limit (~1.5-2 eV) due to the suppression of interlayer interactions [4,5]. This strong variation in the electronic structure allows for optical identification of variations in layer thickness [6,7], which, in addition to the high carrier mobility, make black phosphorus extremely promising for device applications based on 2D layered materials [8-13]. Furthermore, black phosphorous contains an additional distinction: a strongly anisotropic band structure along the principle direction Γ-X compared to Γ-Y, which can be characterized by two free electron-like bands with starkly different effective masses [14,15]. To date, measurable anisotropies have been reported for in-plane carrier mobility, optical extinction, thermal transport, and plasmon dispersion [14,16-19].

In addition to device applications, it has been proposed that the band structure of black phosphorus can be manipulated by strain or electric fields toward a superconducting phase transition [20,21] or a Lifshitz topological transition [15,22-24]. The first experimental work to this end used surface alkali metals to controllably close the band gap and demonstrate a semi-metal transition in which one of the two free electron bands transformed into a Dirac-like band [25,26]. This opens up the possibility of controlling both the topology and band gap of black phosphorus based on surface electric fields.

However, unlike graphene, $MoS_2$, or $WS_2$, black phosphorus is extremely reactive, making its electronic properties sensitive to ambient conditions [27,28]. This motivates a careful understanding of the intrinsic doping of black phosphorus and how native defects modify the band structure [29-32]. Scanning tunneling microscopy/spectroscopy (STM/STS) has been used to characterize the intrinsic band gap of bulk black phosphorus in ultra-high vacuum, but conflicting values have been reported [33, 34]. Here, we characterize the structural and electronic properties of black phosphorus and the influence of inherent single vacancies on those properties, utilizing a combination of low temperature STM/STS with electronic structure calculations. We illustrate that black phosphorus is *p*-doped with a band gap of ~0.3 eV; this doping is facilitated by single vacancies which are characterized by an anisotropic charge density and in-gap resonance states near the valence band edge. Combining tight-binding calculations with atomic-scale STM/STS, we identify the structure of these vacancies as well depth profile them, correlating their electronic properties near the band gap with their depth from the surface. Moreover, we demonstrate via quasiparticle interference that these vacancies act as strong scattering centers illuminating the anisotropic band structure.

Black phosphorus cleaves along the [001] direction due to weak van der Waals interactions; the STM image in Fig. 1a shows the resultant atomic structure after cleaving *in-situ*, which leaves a macroscopically flat black phosphorus (001) surface. Constant-current STM images reveal zig-zag rows (along the [010] or zig-zag direction), which correspond to the upper plane of phosphorus atoms (light orange atoms in Fig. 1b) in the topmost layer of the buckled rhombohedral structure. As each phosphorus atom contains a lone pair of non-bonding electrons, the charge density per atom extends quite far into the vacuum, leading to the significant (0.02nm) atomic corrugation observed in the STM topography. The crystal structure yields two mirror-symmetric sublattices, which we label A and B in Fig. 1b, within each [010]-oriented atomic row. In general, both sublattice sites appear equivalent in STM (Fig. 1c), except in the vicinity of a defect. A close examination of the atomic structure (Fig. 1c) enables the determination of the lattice parameters within the (001) plane: $a = 0.32$ nm and $c = 0.44$ nm. These parameters are consistent with previous experimental and theoretical studies of unstrained black phosphorus [1, 35]. A FFT of the atomic resolved image in Fig. 1a, (Fig. 1d) allows for the identification of the high-symmetry points of the topmost layer, where we adopt the convention denoting the real-space [100] direction as armchair and [010] as zig-zag.

To understand the electronic properties of black phosphorus, STS was performed to characterize the band gap. Using the analysis shown in Fig. S1 for $n = 48$ individual measurements, we estimate $E_g = 0.317 \pm 0.036$ eV. This band gap is consistent with previous large-area measurements on bulk black phosphorus crystals (0.31-0.35 eV) [1]. Scanning tunneling spectroscopy further illustrates that the Fermi level ($E_F$) generally resides within ± 30 mV of the valence band edge (Fig. S2); considering the band gap, this indicates that black phosphorus is strongly *p*-doped without deliberate modification. To elucidate the origin of this intrinsic *p*-doping, we investigate below the structural and electronic properties of the atomic-scale defects in black phosphorus.

Figure 2a shows a representative large-area constant-current image of black phosphorus near the valence band edge. The surface is decorated with a low density of characteristic, elongated, dumbbell-shaped protrusions, which exhibit enhanced charge density along the [100] (armchair) direction. As these features are present after cleaving in ultra-high vacuum, we associate them with native defects in the black phosphorus crystal. Line profiles across these defects, shown in the upper left inset of Fig. 2a, reveal that measured heights of the defects range from 0.01 nm to 0.2 nm above the pristine surface, spanning nearly an order of magnitude greater than the intrinsic corrugation of the atomic lattice. The monotonic changes in apparent height (Fig. 3) are consistent with STM studies of defects at various heights beneath the surface of III-V semiconductors [36-39], as well as in various layered compounds [40, 41]. We discuss the depth profiling of the defects in more detail later.

Atomic-scale characterization of the observed dumbbell structures is shown in Fig. 2b - c, for two different depths below the surface. Close examination of these images reveals that the two dumbbell structures can be transposed onto one another through a reflection along [010] (see also Fig. S3 – S4). Utilizing tight-binding calculations, we investigate the resultant charge density associated with midgap states near single vacancies (Fig. 2d - e) in the uppermost layer at both sublattice positions (A, B). The calculations qualitatively reproduce the highly anisotropic charge density seen in STM images (Fig. S5), as well as a sublattice-dependent orientation of the charge density. By comparing the curvature near the defect center, each experimentally observed dumbbell structure can be directly assigned to a vacancy at a particular sublattice. Therefore, we refer to the vacancies at sublattice A and B, as vacancy A and B, respectively. Furthermore, tight-binding calculations show that vacancies below the first layer maintain the dumbbell defect shape, but have diminished intensity with respect to vacancies in the surface layer (Fig. S6). In reproducing key experimental observations, tight-binding calculations reveal that it is possible to trace each observed defect to a particular depth below the surface as well as the atomic sublattice within that layer.

In order to thoroughly understand the relationship between the experimentally determined apparent height and the vacancy depth below the surface, over 100 vacancies were characterized according to their sublattice location and constant-current height profile. This analysis, with the index sorted according to apparent height, is presented in Fig. 3 for both vacancy types. From the total number of characterized vacancies, we observed vacancy B nearly twice as frequently as vacancy A. The plots in Fig. 3 display the mean intensity extracted from line profiles of the apparent height (illustrated in the insets) for (1) both sides of the vacancy (red/blue), (2) the left side (gray square), (3) and the right side (black diamond). Clear trends emerge in the apparent height for vacancy A, seen in the step-like features at approximately 0.01 nm, 0.04 nm, 0.07 nm, and 0.10 nm. The plateaus are attributed to vacancies residing in discrete layers below the surface, which we label by increasing number with increasing depth underneath the surface, with the label 0 referring to vacancies in the surface layer. Furthermore, a majority of A vacancies exhibited an enhanced intensity on the left region of the charge density compared to the right, with asymmetry between regions decreasing with increasing depth below the surface. Height plateaus for vacancy B are not readily identifiable in Fig. 3b; however, there is a height dependent trend in the d$I$/d$V$ spectra, as we show below (Fig. 4a and Fig. S6), similar to vacancy A, which we believe is related to the specific layer assignment. The preference of vacancy B compared to vacancy A, combined with the differences in their height profiles indicates that the symmetry of the A-B sublattices is broken near the surface at the vacancy site. We attribute this trend to surface relaxation effects, which may relax the two sublattices asymmetrically at the vacancy site near the surface. While the symmetry breaking of the A-B sublattices may be due to vertical relaxation, we note that we cannot rule out a small vertical relaxation for deeper vacancies, which show A-B sublattice symmetry.

To correlate the electronic properties of each vacancy type with their depth from the surface, spatially-resolved tunneling spectroscopy was performed for vacancies at various depths from the surface (Fig 4). All spectra for both vacancy types taken near the center of the vacancy share two key departures from the pristine band gap spectrum (gray curve at the bottom of Fig. 4a,b): (a) spectral resonance(s) emerge in all curves near $E_F$ and (b) non-zero tunneling conductance is measured throughout most of the bulk band gap in the proximity of the vacancy (see Fig. S7). Moreover, the spectra (Fig. 4a and 4b) clearly show a phenomenological trend with respect to the resonance states near $E_F$. The upper curves ($h > 0.06$ nm) show two distinct spectral features split (not necessarily equivalently) about the Fermi level. For both vacancy types, these split peaks are seen for vacancies near the surface, whereas deeper vacancies converge to a single spectral resonance slightly above $E_F$ (see also Fig. S8). Furthermore, while the absolute energy of the double-resonance structure for vacancy B varies, the separation between the resonance features is nearly constant with a value of approximately $27 \pm 3$ mV. The rigid separation likely originates from the monovalent nature of single vacancies, while the absolute position of spectral features can be influenced by the depth from the surface and the surrounding environment. Additionally, the relative intensity of the spectral peaks varies with the precise measurement position (see Fig. S9). It is important to note that we do not observe any significant effects of tip-induced band bending (Fig. S10), most likely due to the pinning of the valence band. At $h < 0.05$ nm, the spectra for vacancy A and B merge to a single resonance at $9 \pm 2$ mV. This further indicates that the symmetry between both sublattices may be broken by surface relaxation for vacancies at or near the surface (e.g. 0L-2L), and this symmetry is restored in the bulk. These spectral features reveal that the atomic vacancies host shallow donor states, which contribute to the inherent $p$-doping and lead to electrical conduction within the band gap.

Spatially resolved STS can also be utilized to map the lateral variation of the observed spectral features. The line spectra in Fig. 4c show that the spectral resonances near the Fermi level are localized to the topographically bright dumbbell regions (red point in inset and red curve, see also Fig. S7). Such resonance states are observable as far as 5 nm from the vacancy site along [100]. In the direction of [010] (Fig. 4d), however, the resonance states decay much faster, generally within 2 nm of the vacancy site. The strong wavefunction anisotropy likely stems from the drastic differences in the effective masses of the valence band along Γ-X and Γ-Y. The full spatial profile of the differential tunneling conductance is shown in Fig. S7, at energies both off and on resonance.

The anisotropic nature of the black phosphorus band structure is strongly pronounced in quasiparticle interference (QPI) near the vacancy sites [42, 43]. Figure 5a shows a constant-current d$I$/d$V$ image acquired at $V_S$ = -0.6 V near several vacancies with varying depth from the surface. Clear oscillations are observed around each vacancy, with a much stronger intensity along [100] compared to [010]. This asymmetric scattering pattern is a signature of the anisotropy of the valence bands near the Fermi edge, giving rise to nesting of the quasiparticle scattering vectors. This asymmetry can be clearly traced in the Fourier transform of the d$I$/d$V$ map (Fig. 5b), where higher intensity is seen for vectors oriented along the Γ-X direction. The edge of the Brillouin zone (BZ) is also highlighted with a green rectangle for reference. As seen in the calculated constant energy contours in Fig. 5d (progressive ellipses around Γ, see Fig. S9 for cuts on the band structure), the shape of the valence band is strongly ellipsoidal near $E_F$. This ellipticity causes an imbalance of available scattering vectors in the Γ-X versus Γ-Y directions leading to more prominent carrier scattering along [100]. As the calculations reveal, the constant energy contours become more circular at increasing negative energies, which weakens the nesting condition and results in stronger intensity fringes along [010], as shown in Fig. 5c.

In conclusion, we have characterized the structural and electronic properties of inherent single vacancies in black phosphorus. We illustrate that vacancies exhibit a highly anisotropic and delocalized charge density. Moreover, intrinsic vacancies possess in-gap resonance states near the valence band edge, which strongly perturb the observed bulk band gap. Therefore, we conclude that intrinsic vacancies contribute to the strong *p*-doping of black phosphorus. This work motivates further studies of the bulk doping, to conclude if intrinsic vacancies are solely responsible for the observed strong *p*-doping. We further reveal that these single vacancies serve as strong scattering centers, as seen in quasiparticle interference near the observed vacancy sites. This study provides fundamental insight into the influence of intrinsic point defects on the electronic properties of black phosphorus, which is crucial for future developments of black phosphorus technologies. Moreover, the recent findings of magnetic vacancies in graphene and indications of magnetism in black phosphorus vacancies [32], make this an interesting platform to study magnetic ordering in *sp*-driven systems [44].

**Experimental Methods**

STM/STS measurements were carried out on a commercial Omicron low-temperature STM with a base temperature of 4.6 K, operating in ultra-high vacuum (< 1e-10 mbar), with the bias applied to the sample. Etched W tips were utilized for these measurements, and were treated *in-situ* by electron bombardment, field emission, as well as dipped and characterized on clean Au surface. Scanning tunneling spectroscopy was performed using a lock-in technique to directly measure d$I$/d$V$, with a modulation frequency of $f_{mod}$ = 4.2 kHz and amplitude of $V_{mod}$ = 5 mV. Black phosphorus crystals were provided by HQ graphene and stored in vacuum at a temperature less than 25 °C, cleaved under ultra-high vacuum conditions at pressures below 1x10$^{-9}$ mbar, and immediately transferred to the microscope for *in-situ* characterization.

**Theoretical Calculations**

Theoretical calculations were based on the tight-binding model proposed for multilayer BP in Ref. 5. To model structural defects, we considered a missing atom in a three-layer supercell with dimensions (16a x 12c) ~ (52.5 x 53.0) Å, and performed exact diagonalization of the tight-binding Hamiltonian. Wavefunctions of a defect state were represented as $\Psi(r) = \Sigma_i c_i \cdot \varphi(r)$, where $c_i$ is the contribution of the i-th atom given by the corresponding eigenvector of the Hamiltonian, and $\varphi(r)$ is the cubic harmonic representing the $3p_z$-like orbital of P atoms, being the basis functions of the Hamiltonian. Constant energy contours appearing in Fig. 5d were extracted directly from first-principles GW calculations performed in Ref. 5.

The authors declare no competing financial interests.

Additional information includes band gap determination and statistics, detailed examination of defect anisotropy, extended tight-binding results, spatial profiles of defect-related in-gap conduction, representative depth dependent d$I$/d$V$ spectra, spatial variation in d$I$/d$V$ spectra, height dependent d$I$/d$V$ spectra, and the calculated band structure of black phosphorus can be found in the supplementary material. This material is available free of charge from the internet at http://pubs.acs.org.

**Acknowledgments**

B.K., N.H., and A.A.K. acknowledge financial support from the Emmy Noether Program (KH324/1-1) via the Deutsche Forschungsgemeinschaft, and the Foundation of Fundamental Research on Matter (FOM), which is part of The Netherlands Organization for Scientific Research (NWO). B.K., N.H., and A.A.K. also acknowledge the VIDI project: 'Manipulating the interplay between superconductivity and chiral magnetism at the single atom level' with project number 680-47-534 which is financed by NWO. N.H. and A.A.K. also acknowledge support from the Alexander von Humboldt foundation via the Feodor Lynen Research Fellowship. A.N.R. and M.I.K. acknowledge support from the European Union's Horizon 2020 Programme under Grant No.696656 Graphene Core1. A.A.K. would also like to acknowledge scientific discussions with Alexander Grueneis and Daniel Wegner.

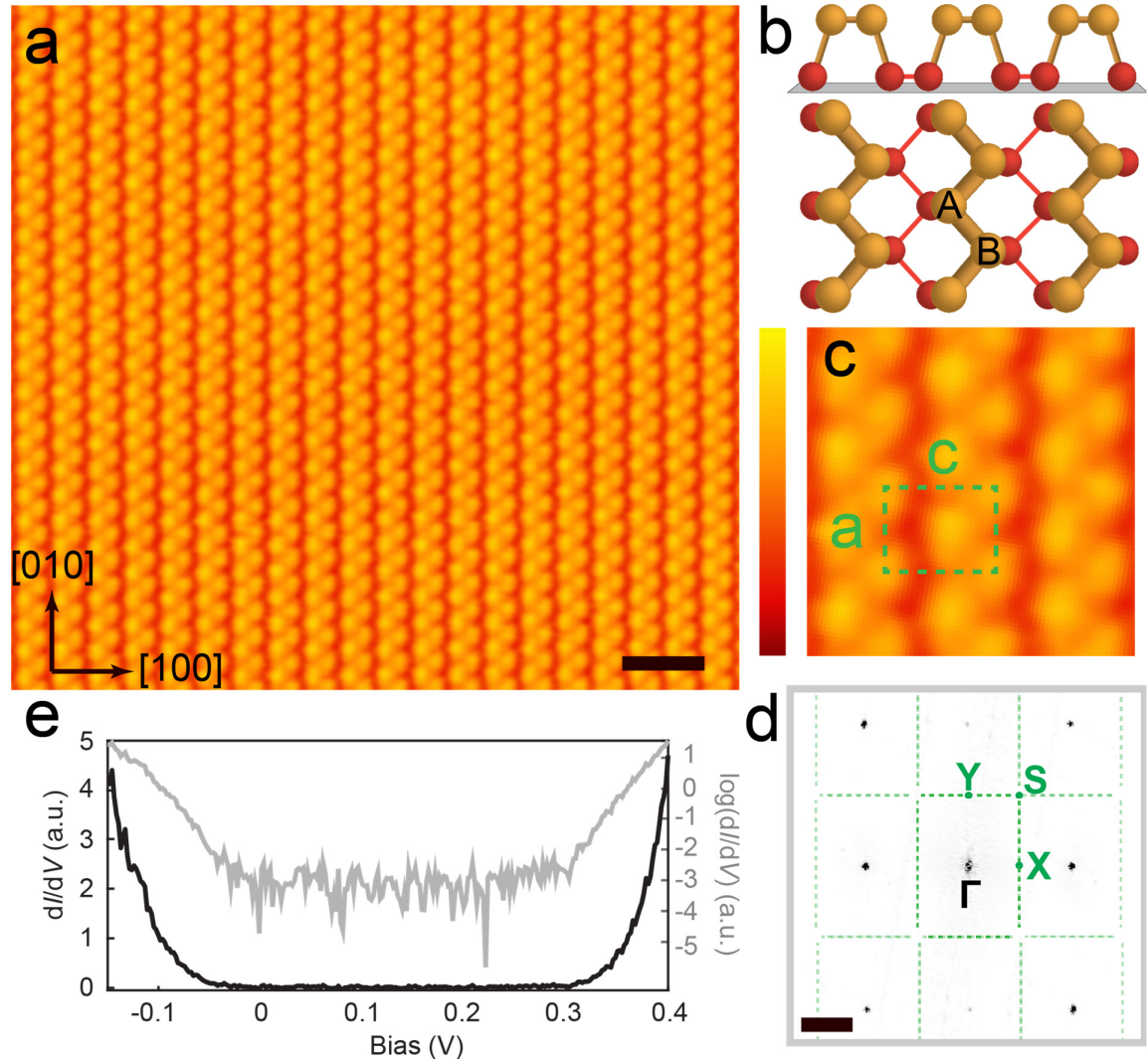

Figure 1. (a) Constant-current STM image of black phosphorus after *in-situ* cleaving ($V_S = -0.1$ V, $I_t = 200$ pA, scale bar = 1 nm). (b) Crystal structure of black phosphorus with side view (upper) illustrating the buckling of each layer. The top view (lower) illustrates the two sublattices (A, B), in comparison to which atoms are imaged with STM (bright orange). (c) Zoomed-in image of black phosphorus, with lattice parameters $a = 0.32$ nm and $c = 0.44$ nm (color bar: 0 to 0.02 nm). (d) Fourier transform of the STM image in (a) with green dashed lines showing Brillouin zone boundaries and the principle in-plane reciprocal space locations denoted (scale bar = $\pi/c$ = 7.14 nm$^{-1}$). (e) STS of black phosphorus ($V_{stab} = 0.35$ V, $I_{stab} = 100$ pA, $V_{mod} = 5$ mV, $f_{mod} = 4.2$ kHz) illustrating an electronic band gap of 0.317 +/- 0.036 eV. The logarithm of the d$I$/d$V$ tunneling spectra is shown in gray to accentuate the noise floor within the gapped region. The band gap was derived from n = 48 independent spectra.

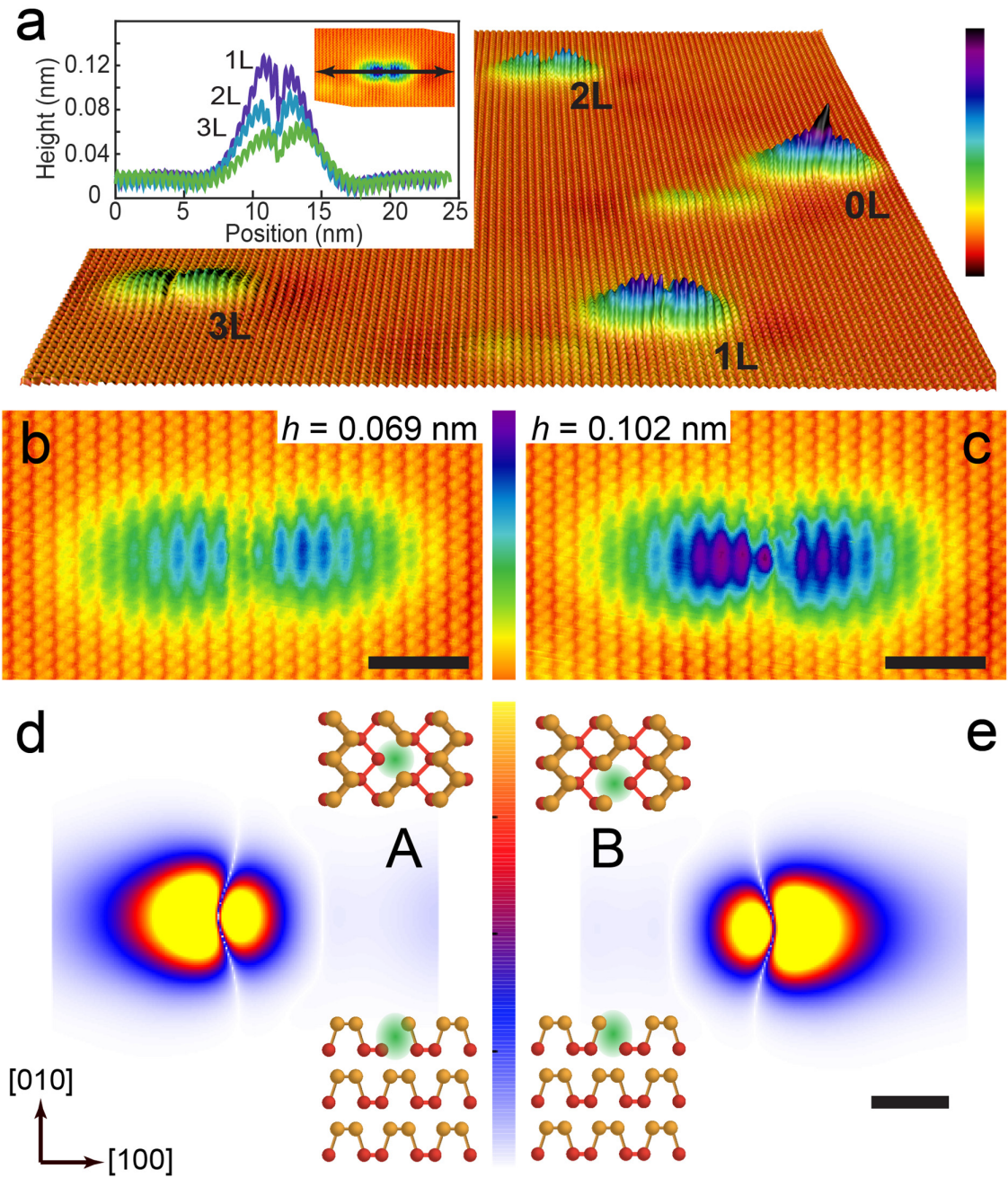

Figure 2. (a) 3D representation of a constant-current STM image with a distribution of single vacancies in black phosphorus ($V_S$ = -0.1 V, $I_t$ = 200 pA, size: 62 nm x 48 nm, color bar: 0-0.2 nm). (Inset) Line profiles taken across the vacancies labeled 1L, 2L, and 3L in (a). Constant-current STM image of a single vacancy at (b) sublattice A, and (c) at sublattice B ($V_S$ = -0.1 V, $I_t$ = 200 pA, scale bar = 2 nm). (d) Tight-binding calculations of the charge density of a single vacancy in black phosphorus located at (d) sublattice site A and (e) sublattice site B (scale bar = 1 nm).

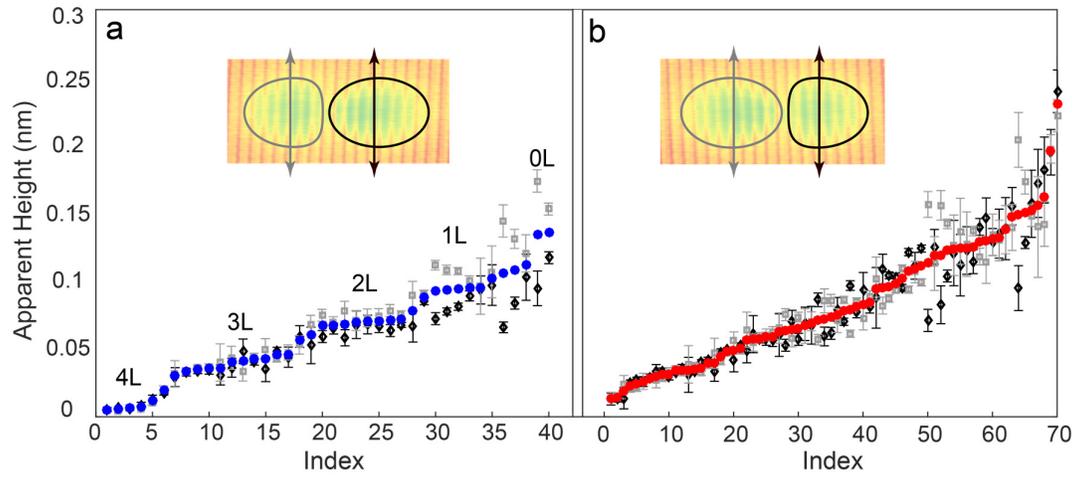

Figure 3. (a) Scatter plot of the averaged apparent STM height for vacancy A (see inset). Blue points refer to the mean intensity from both sides of the vacancy (black and gray contours in inset), while the gray squares correspond to the left side with estimated error and the black diamond shows the height of the right side with corresponding error. (b) Scatter plot of the average apparent height for vacancy B. Red points refer to the mean intensity from both sides of the vacancy, while the gray square shows the left side with estimated error and the black diamond shows the height of the right side with corresponding error. For all data points: $V_S$ = -0.04 V, -0.05 V, or -0.1 V, $I_t$ = 100-200 pA.

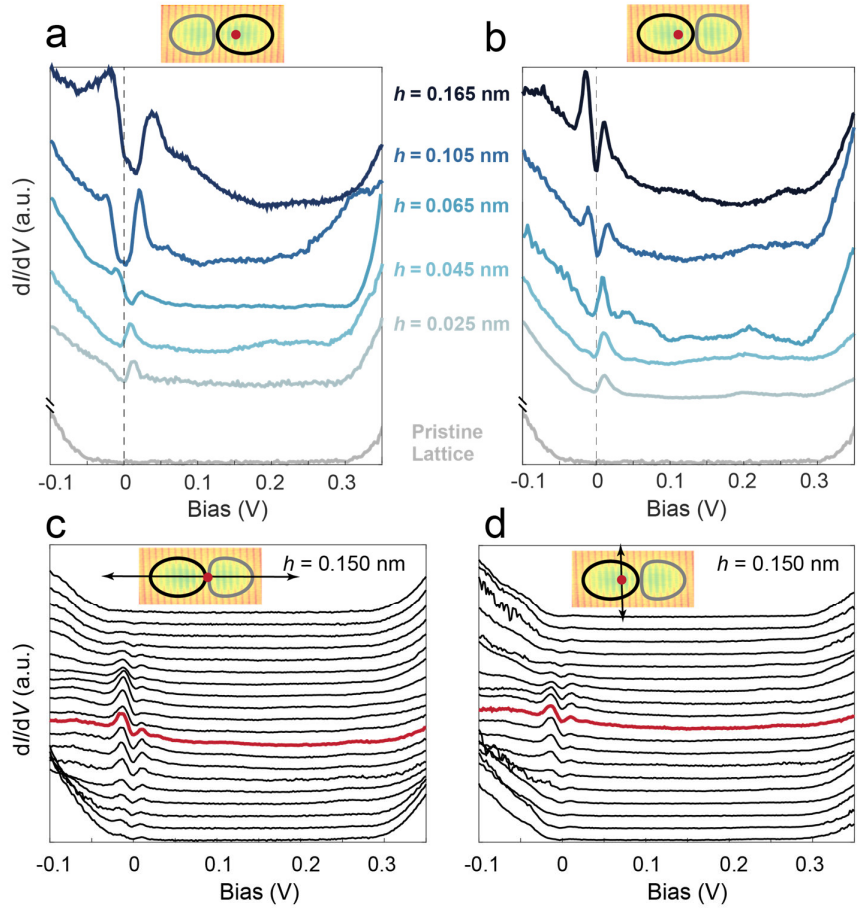

Figure 4. Tunneling spectroscopy of the depth dependence and spatial variation for both vacancy types taken at red circles. (a-b) Representative tunneling spectra sorted by decreasing apparent height ($h$) according to progressive blue scale in the middle for (a) vacancy A, and (b) vacancy B taken at the position of the red dot as denoted in the inset. All apparent heights are within ± 0.01 nm, except for the $h$ = 0.165 nm, which is ± 0.02 nm. (c-d) Spatially resolved tunneling spectra taken for vacancy B ($h$ = 0.150 nm) along: (c) [100] (armchair) direction over a total length of 20 nm and (d) along [010] (zig-zag) direction over a total length of 10 nm. Stabilization conditions for all spectra: $V_s$ = 0.35 V, $I_t$ = 200-400 pA. The red spectra were taken at the red point in the inset.

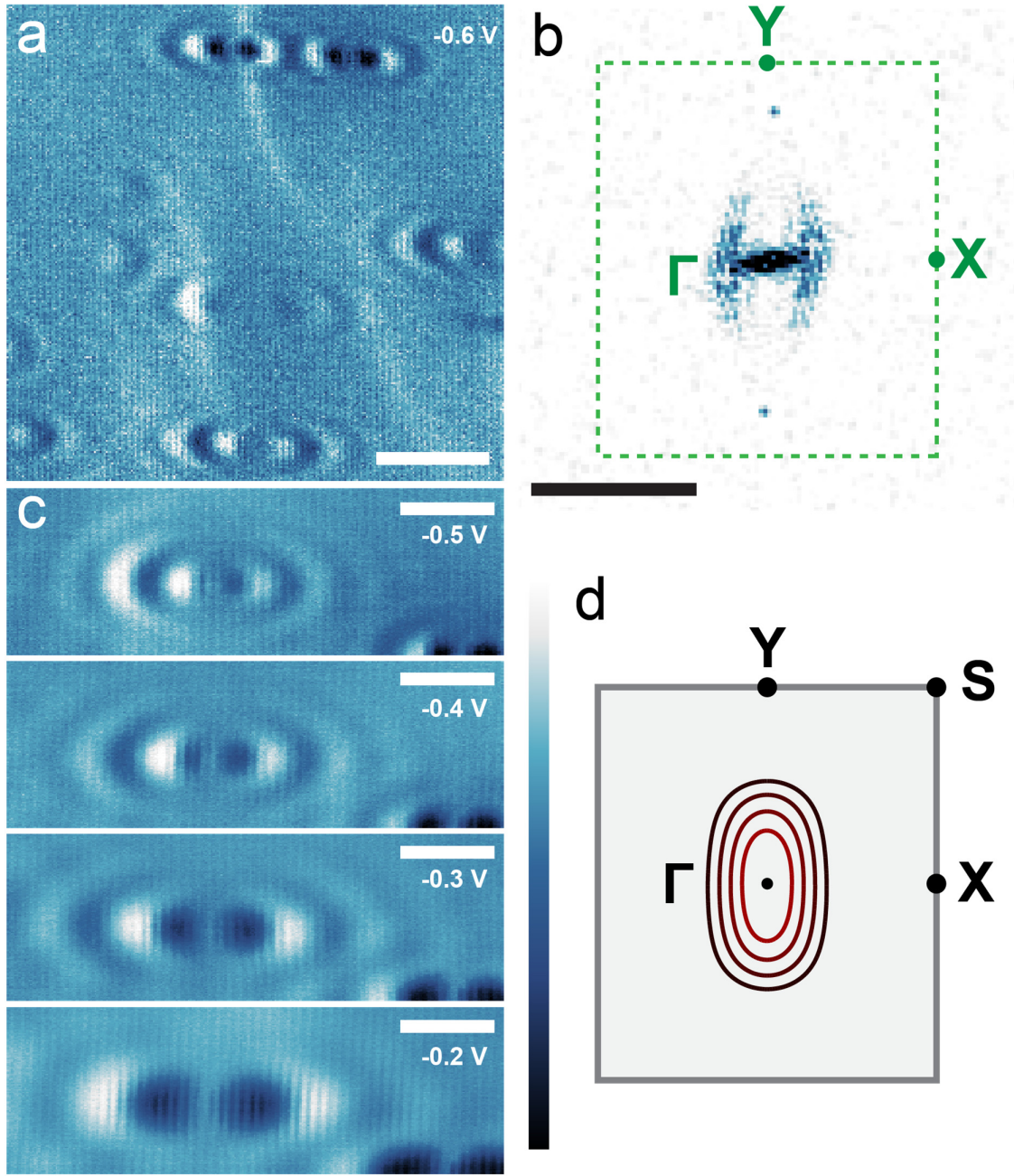

Figure 5. (a) STS map (taken concurrently with constant-current topography) at $V_s$ = -0.6 V showing interference patterns surrounding point defects in BP ($V_s$ = -0.6 V, $I_t$ = 600 pA, scale bar = 8 nm). (b) FFT of STS image in (a) (scale bar = $\pi/a$ = 7.14 nm$^{-1}$). The open-ended ellipse around Γ corresponds to valence band carrier scattering at defect sites. (c) Dispersion of scattering behavior at indicated energies ($V_s$ = varying, $I_t$ = 600 pA, scale bars = 4 nm). (d) Theoretical calculation of valence band constant energy contours in the vicinity of Γ, shown with the BZ for comparison. From dark to light red (outer to inner): $E$ = -0.9 eV, -0.7 eV, -0.5 eV, -0.3 eV. For (a) and (c) $V_{mod}$ = 5 mV and $f_{mod}$ = 4.2 kHz.

**Supporting Information**

# Probing Single Vacancies in Black Phosphorus at the Atomic Level

Brian Kiraly, Nadine Hauptmann, Alexander N. Rudenko, Mikhail I. Katsnelson, Alexander A. Khajetoorians[*]

Institute for Molecules and Materials, Radboud University, 6525 AJ Nijmegen, Netherlands

[*] Correspondence to: a.khajetoorians@science.ru.nl

**Band Gap Determination**

The band gap values presented in Fig. 1a and Fig. S2 were determined using the following protocol (Fig. S1), slightly modified from Ugeda *et.al* [1]. First, the raw d$I$/d$V$ curves were vertically offset by a factor of 1.1 times the minimum conductance value to ensure no data points were negative. Second, the logarithm of the data was taken to delineate the noise floor of these measurements. Using the log data, the in-gap conductance $C_{g,av}$ was determined within the range of 0.0 - 0.2 eV. The standard deviation $\sigma_{g,av}$ about $C_{g,av}$ was also determined. Due to non-linearities in the band onsets, the band edges were determined by the simple condition log(d$I$/d$V$) > $C_{g,av} + \sigma_{g,av}$. To ensure that no spurious points fulfilled the necessary band gap condition (due to larger than normal in-gap noise for instance), the data have been smoothed with a 3-point moving filter for the thresholding step. The band gap data for three different threshold conditions (referred to as $0\sigma_{g,av}$, $1\sigma_{g,av}$, and $2\sigma_{g,av}$) is shown in Fig. S1 to illustrate the variation of the extracted band gap; in general, with spectra demonstrating sharp band onsets (48 spectra in total), the difference between the $1\sigma_{g,av}$ and $2\sigma_{g,av}$ data is less than 10%. In Fig. S2, the error bars are derived from the difference between the $1\sigma_{g,av}$ and $2\sigma_{g,av}$ band edges.

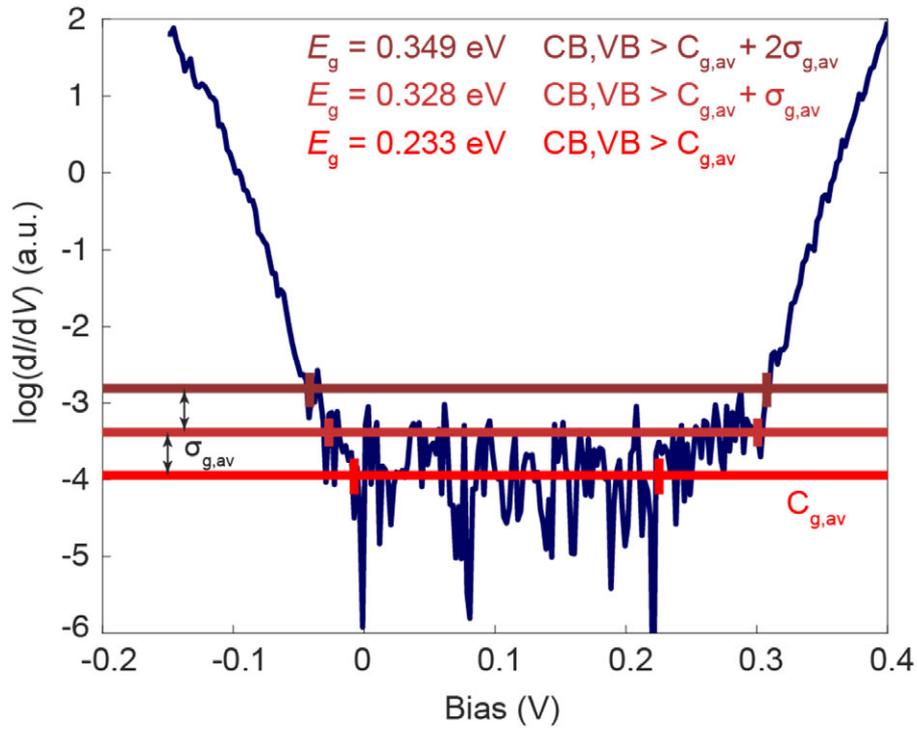

Figure S1. Logarithm of a single d$I$/d$V$ band gap spectrum of bulk black phosphorus and the procedure used for determining the band gap from tunneling spectroscopy. The horizontal lines indicate the average conductance within the band gap ($C_{g,av}$) (red) and the $C_{g,av} + \sigma_{g,av}$ (darker red) and $C_{g,av} + 2\sigma_{g,av}$ (darkest red) levels used for the band gap determination.

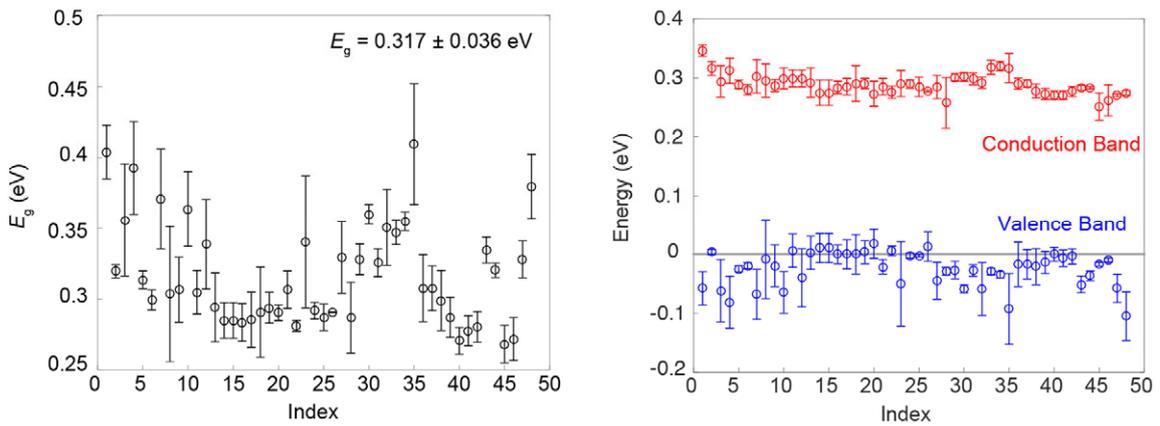

Figure S2. (Left) Extracted band gap values with error using the procedure outlined in Fig. S1 for 48 distinct measurements. (Right) Position of valence band maximum and conduction band minimum for the each individual measurement on the left. Gray horizontal line shows the location of the Fermi level.

**Detailed Examination of Vacancy Anisotropy**

Fig. S3a shows a 2D view of the 3D figure from Fig. 2a in order to more clearly illustrate the spatial distribution of each vacancy's charge density and the asymmetry in the dumbbell structures. The image contains several vacancies found on both sublattices A and B. In order to more easily illustrate the charge density of each vacancy, we filter out the atomic lattice in Fig. S3b – d. From these images, it is apparent that the two lobes for each vacancy are inequivalent, as near the center one lobe generally exhibits a flattened edge while the other lobe remains curved. The orientation of the flat and curved lobes is mirrored for sublattice A versus sublattice B vacancies. A high-resolution STM constant-current image of a single vacancy is shown in Fig. S4a. This image is also presented in Fig. S4b, with the atomic lattice filtered, to further illustrate the asymmetry of the charge density. The vacancy A is clearly distinguished by the flat edge on the left charge density lobe.

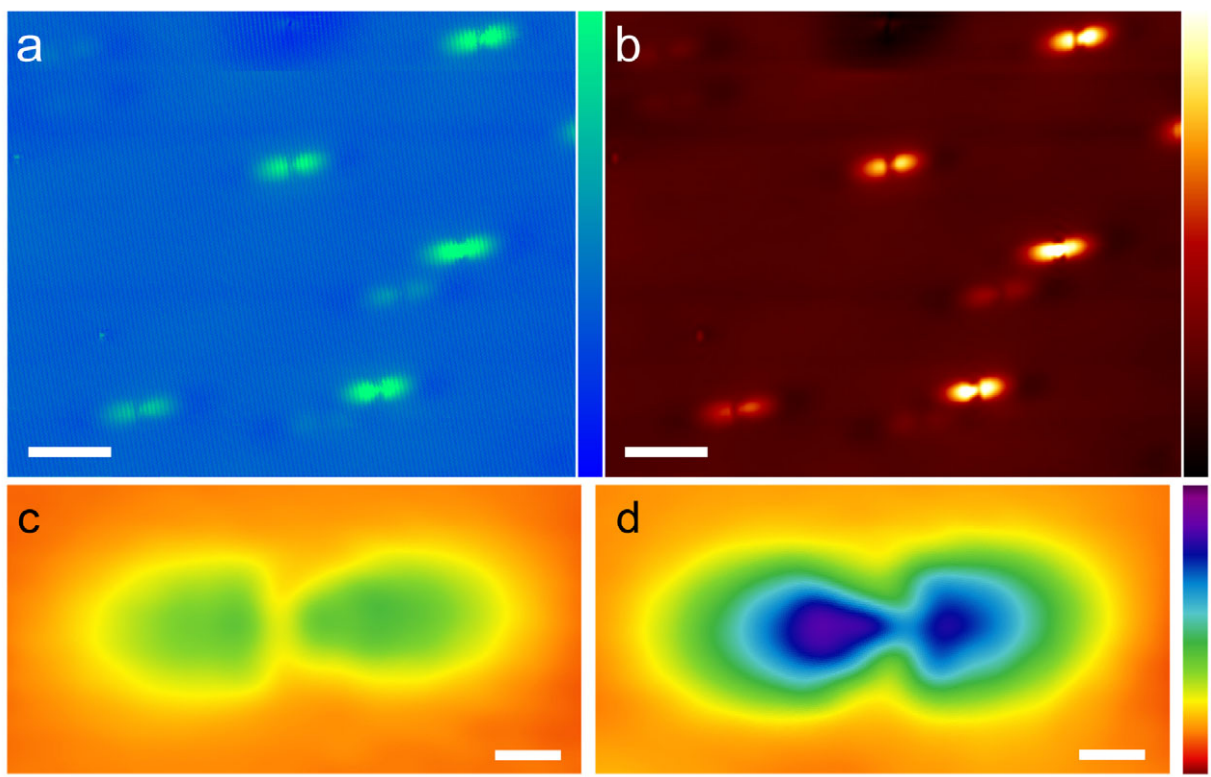

Figure S3. (a) STM constant-current image from Fig. 2a. (b) STM constant-current image from (a) with the atomic lattice filtered out utilizing an FFT filter ($V_S$ = -0.1 V, $I_t$ = 200 pA, scale bar = 8 nm). (c) and (d) show enlarged sections from (b) of vacancy A and vacancy B, respectively ($V_S$ = -0.1 V, $I_t$ = 200 pA, scale bar = 1 nm).

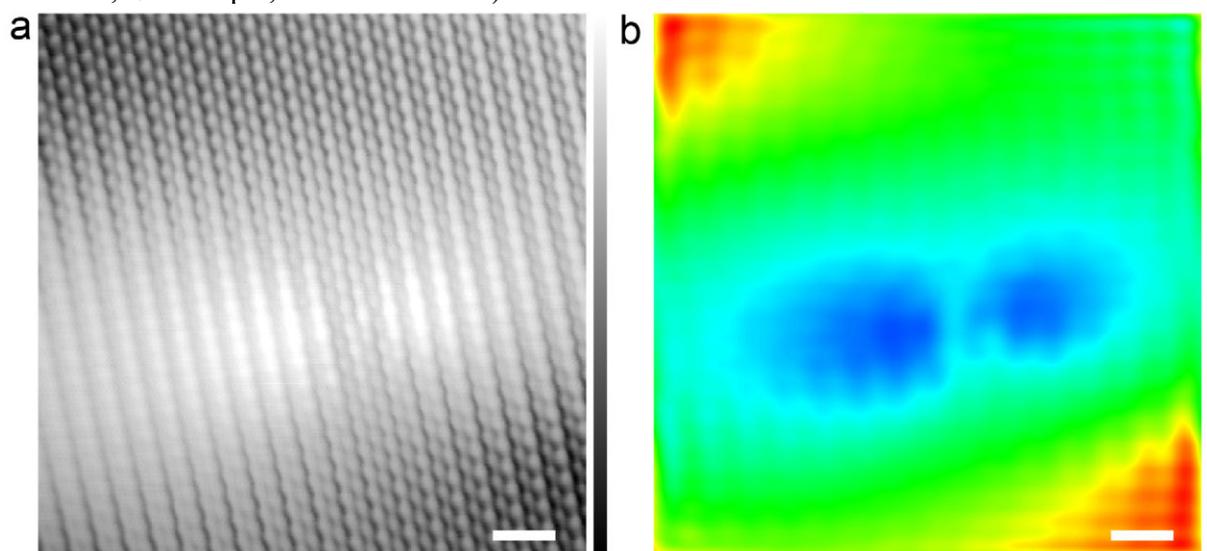

Figure S4. (a) High resolution STM constant-current image of vacancy A. (b) STM constant-current image from (a) with the atomic lattice filtered out utilizing an FFT filter ($V_S$ = 0.008 V, $I_t$ = 100 pA, scale bar = 1 nm).

**Tight-Binding Calculations for Vacancy Wavefunction at Varying Height Above Surface**

Fig. S5 shows a series of wavefunction calculations for vacancy A (top) and vacancy B (bottom) at various calculated heights ($d$) above the BP surface. In order to understand the theoretically predicted structure of buried defects in deeper layers, tight-binding calculations were carried out for vacancies in the surface layer and compared to vacancies in the first phosphorene sublayer. The results are shown in Fig. S6, at selected $d$ values above the topmost phosphorene surface. The images clearly show that the squared wavefunction amplitude at the calculated height is depressed for the lower-lying defects. However, they still retain the anisotropic character of the defects in the first layer, indicating that the varying apparent height is consistent with sub-surface defects.

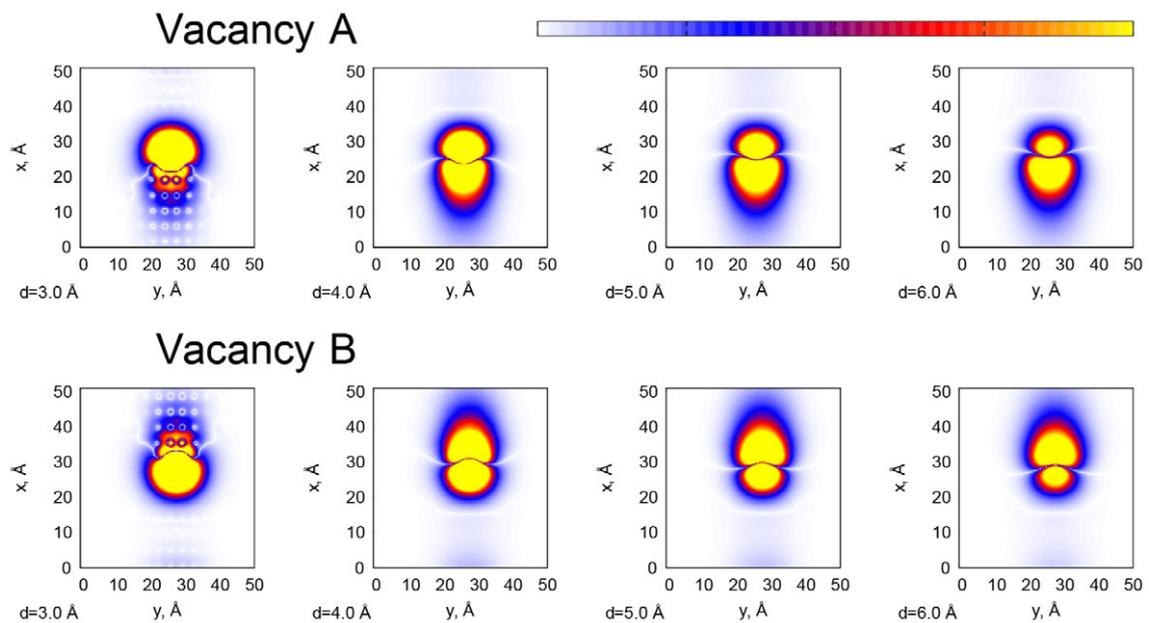

Figure S5. Height dependent square wavefunctions for vacancy A (top) and vacancy B (bottom).

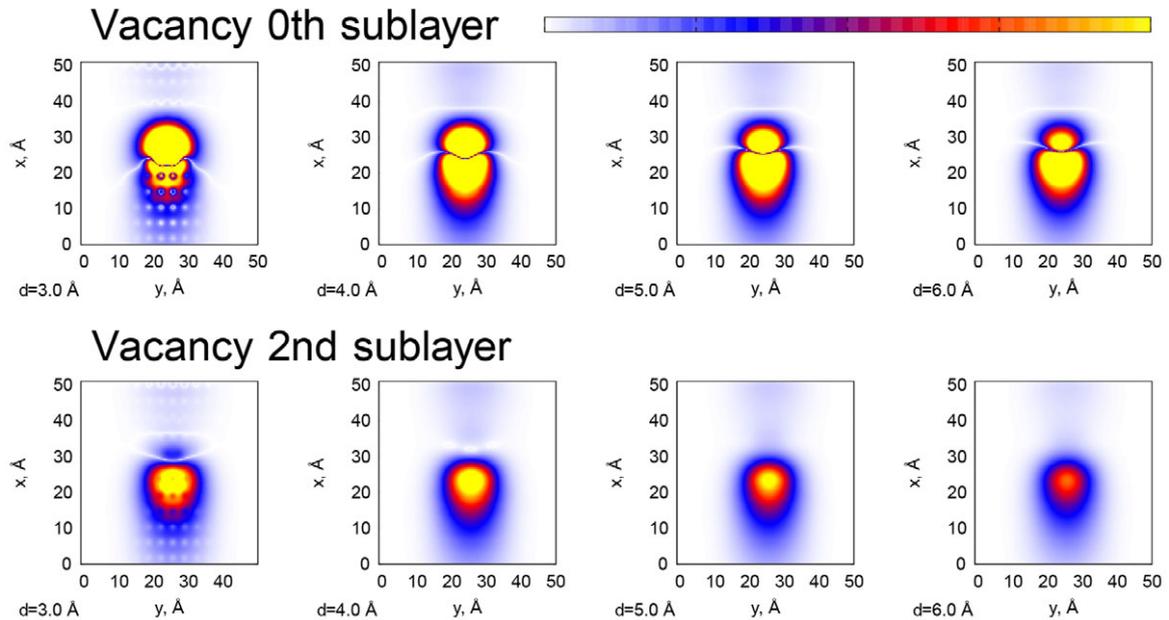

Figure S6. Single defect squared wavefunction for vacancies in first phosphorene layer and subsequent phosphorene layer.

**Spatially Resolved and Layer Dependent d$I$/d$V$ Spectra**

To fully understand the spatial dependence of the spectral resonances and features associated with each defect, spatially-resolved STS measurements were performed. The main results, taken at selected energies to clearly demonstrate the spatial profile of each feature, are presented in Fig. S7. In accordance with previous STS maps, the vacancy exhibits a depressed d$I$/d$V$ signal at voltages outside of the conventional BP band gap. However, as the energy is moved closer to the Fermi level ($V_S$= -0.018 V, -0.011 V, 0.02 V, 0.04 V), in the region where spectral resonances are most pronounced, the d$I$/d$V$ signal is strongly enhanced at the dumbbell structure seen in STM constant-current images. As the energy moves closer to the conduction band edge ($V_S$= 0.06 V, 0.12 V, 0.17 V, 0.24 V), the two-fold symmetry of the dumbbell structure is lost and the enhanced in-gap signal is much more isotropic. Representative, layer-resolved d$I$/d$V$ spectra for defect A and B are shown in Fig. S8. The spectra clearly demonstrate qualitative differences for the two sublattices at $h > 0.06$ nm, but nearly identical features for $h < 0.05$ nm. Fig. S9 shows a series of spectra taken on a sublattice B defect at distinct points along the [100] direction of the defect. The spectra illustrate point-to-point variation in the relative intensities of the spectral features.

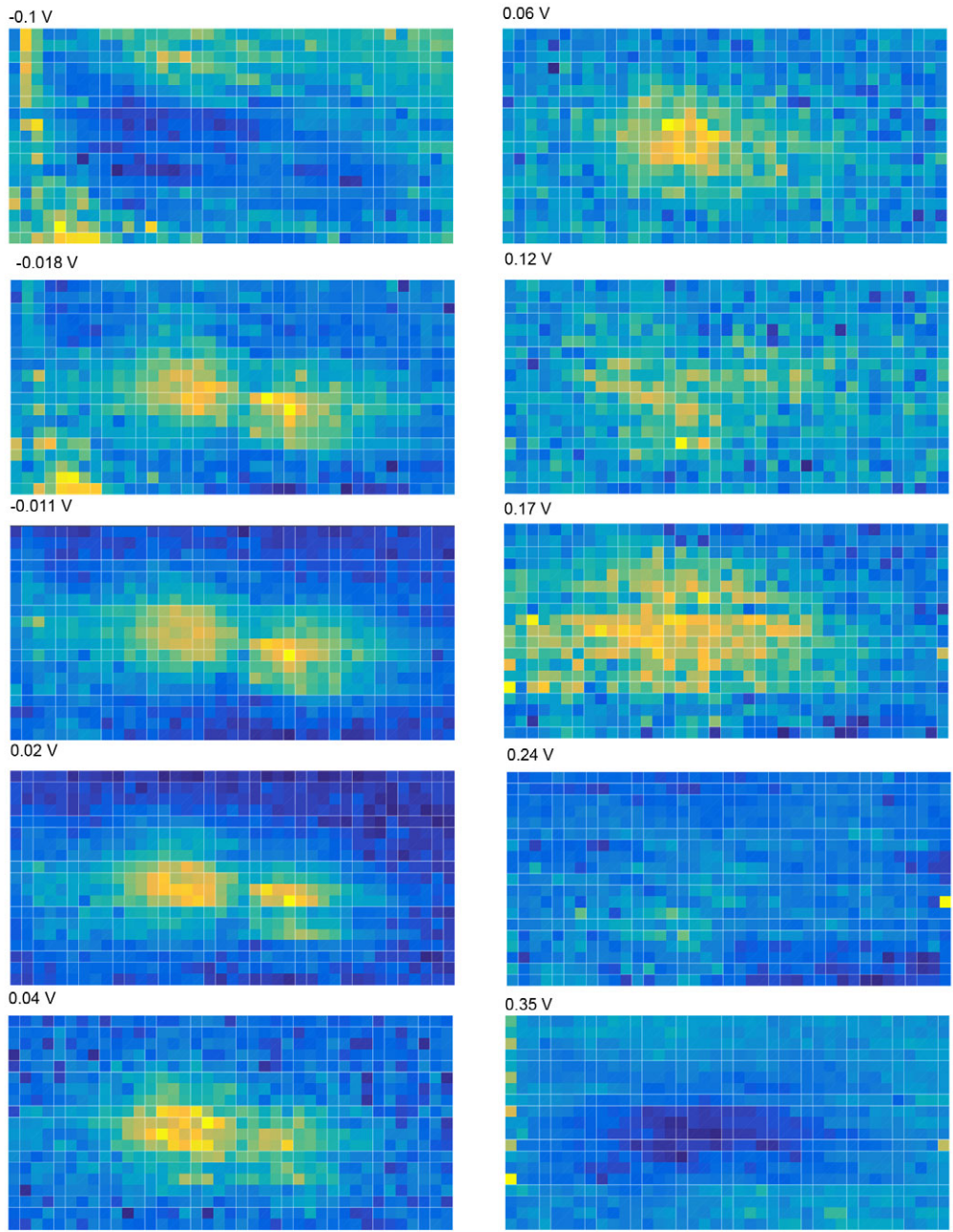

Figure S7. d$I$/d$V$ maps taken at voltages listed above each panel.

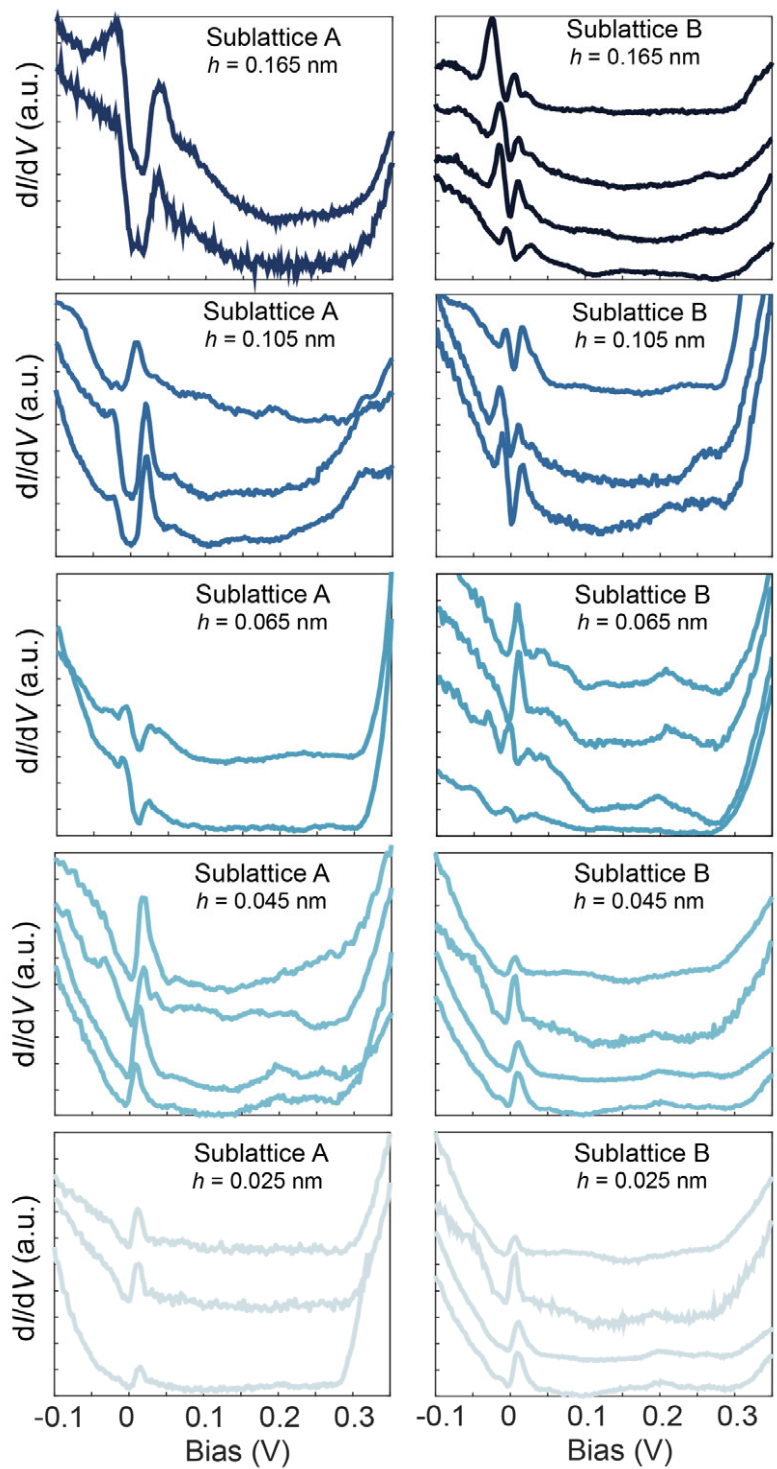

Figure S8. Layer and sublattice-resolved representative point spectra from the BP vacancies. Sublattice A defects are shown on the left, with surface defects at the top (dark blue) and increasing depth further down (light blue). Sublattice B defects are shown on the right, with surface defects at the top (dark blue) and increasing depth further down (light blue). All apparent heights are within ± 0.01 nm, except for the $h$ = 0.165 nm, which is ± 0.02 nm.

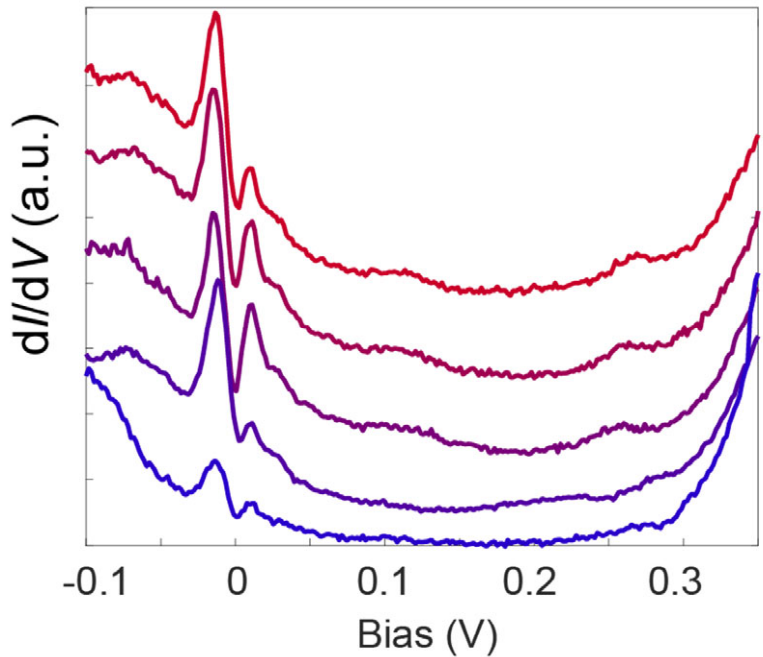

Figure S9. Variation of defect spectra with precise location on the dumbbell defect structure. The curves shown are from vacancy B with $h = 0.150$ nm.

**Tunneling Spectra Dependence on Tip Height**

To determine the impact of tip-induced band bending on the d$I$/d$V$ spectra, we conducted tunneling spectroscopy measurements at increasing tip-sample separations (Fig. S10). To vary the height of the differential conductance measurement, the following $I_{stab}$ conditions were used: 20 pA, 60 pA, 100 pA, 200 pA, 600 pA, with increasing current (decreasing tip-sample separation) corresponding to darker gray. The raw d$I$/d$V$ curves (left) and the log(d$I$/d$V$) curves (right) demonstrate that the band edges are largely unaffected by the modification to tip-sample separation. This indicates that the measured band gaps are insensitive to the electric fields of the STM tip during d$I$/d$V$ measurements.

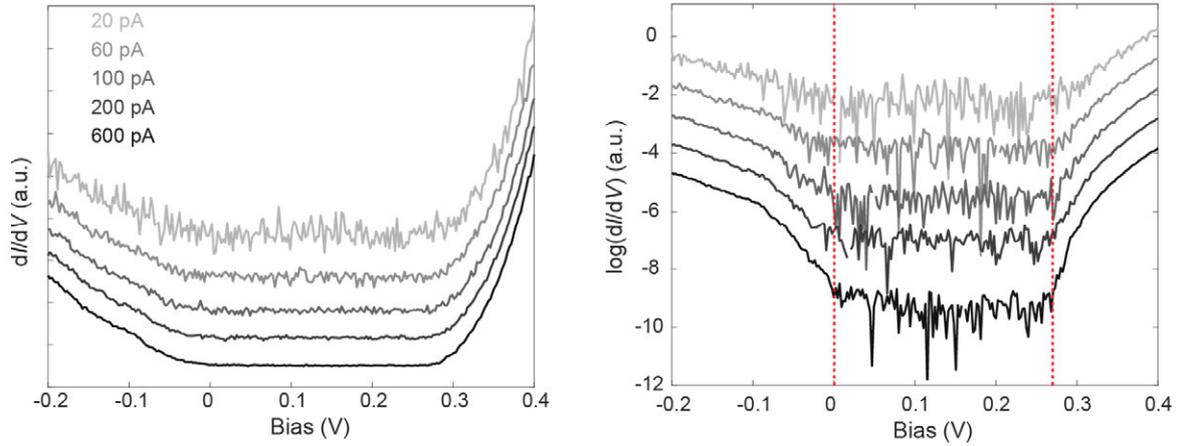

Figure S10. Dependence of the d$I$/d$V$ spectra on the setpoint conditions with the unmodified curves on the left and the logarithm on the right. Starting from the top (light gray): $V_{stab}$ = 0.4 V, $I_{stab}$ = 20 pA, 60 pA, 100 pA, 200 pA, 600 pA. Red lines on the left indicate the position of the electronic band gap as determined from the procedure in fig. S1.

**Iso-Energy Cuts on the Band Structure**

The band structure of bulk BP, calculated under the *GW* approximation [2], is shown in Fig. S11. The energies used for the isoenergy contours shown in Fig. 5d are shown as horizontal dashed lines in Fig. S11 with the same color scale: lighter red to black from -0.1 to -0.9 eV. The band structure shows the origin of the contours' ellipticity: the contrast in dispersion between Γ-X and Γ-Y.

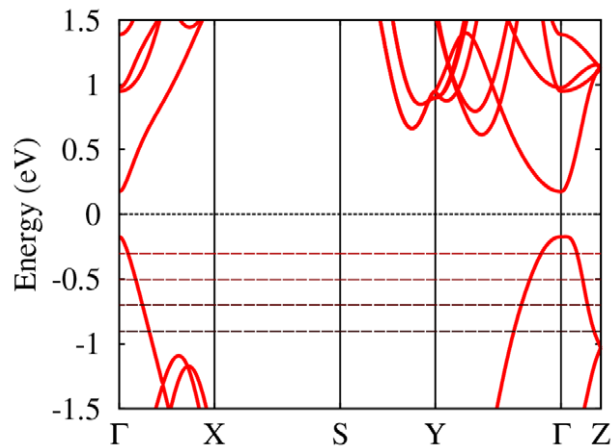

Figure S11. Band structure for bulk black phosphorus with progressive red lines corresponding to the cuts taken for the constant-energy contours in Fig. 5d.